\documentstyle[aps]{revtex}
\begin{document}
\draft
\title{Many Body Diffusion and Interacting Electrons in a Harmonic Confinement}
\author{F. Luczak, F. Brosens \thinspace and J. T. Devreese\thanks{%
Also at Universiteit Antwerpen (RUCA), Groenenborgerlaan 171, B-2020
Antwerpen and Technische Universiteit Eindhoven, NL 5600 MB Eindhoven, The
Netherlands.}}
\address{Departement Natuurkunde, Universiteit Antwerpen (UIA),\\
Universiteitsplein 1, B-2610 Wilrijk}
\author{L. F. Lemmens}
\address{Departement Natuurkunde, Universiteit Antwerpen (RUCA),\\
Groenenborgerlaan 171, B-2020 Antwerpen}
\date{April 20, 1999}
\maketitle

\begin{abstract}
We present numerically exact energy estimates for two-dimensional electrons
in a parabolic confinement. By application of an extension of the recently
introduced many-body diffusion algorithm, the ground-state energies are
simulated very efficiently. The new algorithm relies on partial
antisymmetrization under permutation of particle coordinates. A comparison
is made with earlier theoretical results for that system.
\end{abstract}


\section{Introduction}

In this paper we apply the recently developed quantum Monte Carlo algorithms 
\cite{LBDL98,LBDL98b} to an important model for interacting charge carriers
in quantum dots. The fabrication of novel miniaturized semiconductor
structures on nanometer scale has shed light on a variety of advanced
physical systems and devices in which the classical description of
electronic properties breaks down. The band structure of quantum wells,
multiple quantum wells and superlattices \cite{YU} makes mobile carriers
locate parallel to the semiconductor interfaces and hence induces a
quasi-two-dimensional confinement \cite{CHE93}. Electrons in sandwiched
semiconductor layer structures can be confined perpendicular to the growth
direction \cite{KAS92,KAS93,HEI93}. If the lateral confinement is of the
same order as the electron wavelength, the electrons have essentially no
free direction left. The resulting quasi-zero-dimensional structure is
addressed as a quantum dot or an artificial atom \cite{KAS93}. Mostly, the
confinement effect along the growth direction of the layers is much stronger
than perpendicular to it. Accordingly, on the microscopic scale, the
electrons form two-dimensional disk-shaped objects \cite{HEI93}. Step-like
potential structures, being induced by steps in the conduction band edge,
became a popular object of study in the framework of self-assembled quantum
dots \cite{JAC97}. In field-effect confined quantum dots, experimental
far-infrared transmission spectra indicate the occurrence of
parabolic-shaped confinement potentials \cite{MEU92}. Depending on the gate
voltage applied, the number of electrons may discretely vary from zero to a
large number of electrons \cite{ASH96,KOU97}.

Motivated by the fabrication techniques of quantum dots, we here focus on a
commonly used model of $N$ interacting electrons in a harmonic confinement
potential, 
\begin{equation}
H=\frac{\bar{p}^{2}}{2\mu }+V_{\text{c}}(\bar{r})+V_{\text{int}}(\bar{r})
\label{eq:Hamiltonian}
\end{equation}
where 
\begin{equation}
V_{\text{c}}(\bar{r})=\frac{1}{2}\,\mu \,\omega ^{2}\,\bar{r}^{2}\;,\;V_{%
\text{int}}(\bar{r})=\sum_{i=1}^{N-1}\sum_{j=i}^{N}\frac{e^{2}}{2\,\epsilon
\,\left| \vec{r}_{i}-\vec{r}_{j}\right| }.  \label{firstb}
\end{equation}
Here, $\omega $ denotes the frequency of the harmonic confinement potential
of the dot, $\mu $ the effective mass of the electrons and $\epsilon $ the
dielectric constant of the material. For convenience, we set $\hbar =$ $%
\omega =e=\mu =1$. The only parameter which then enters in (1) is $\gamma
=\epsilon ^{-1}.$ It serves as a parameter to adjust the strength of the
electron-electron repulsion for a given dot radius.

The derivation of the energy eigenfunctions and eigenvalues of (\ref
{eq:Hamiltonian}) turns out to be challenging in cases where many particles
are involved. Even in the classical case, solutions remain elusive, and the
use of analytical approximations or computational methods is indispensable
to reliably predict the energy spectrum of the system. The consideration of
identical particles makes analytical or numerical treatments still more
difficult. Much effort with different methods has been pursued to
investigate the physical properties of quantum dots \cite{JOH95,JEF97}.
Various approaches have been performed using the Thomas-Fermi approximation 
\cite{PIN98}, many-body perturbation theory \cite{MER91}, or the Hartree-
and Hartree-Fock Ansatz \cite{PFA93}. Bearing in mind that exact analytical
approaches are generally restricted either to the limit of non-interacting
quantum-mechanical particles or exclusive cases \cite{TAU93}, some methods,
as e.g. the Pad\'{e} approximants \cite{GON97,GON97b} or the renormalization
perturbation theory \cite{MAT94,ANI98}, attempt to connect these limits.
Exact numerical solutions are possible by diagonalization of the Coulomb
interaction \cite{MER91,PFA93,HAW93,PAL94,PFA95}. Due to their challenging
numerical expense, diagonalization techniques are limited to few-electron
systems. In particular for many-particle systems, for which diagonalization
techniques quickly exceed the capacities of today's computers, Monte Carlo
techniques turn out to be very useful \cite{INTROQM}. Monte Carlo approaches
have been applied to electrons in a parabolic potentials in refs. \cite
{BOL94,BOL96,MAK98}. Apart from their simplicity and flexibility, the power
of Monte Carlo methods lies in the possibility to obtain estimates which
converge to the exact values with a known statistical error. Although the
computational effort needed for Monte Carlo techniques to obtain a given
accuracy does not necessarily grow exponentially with the particle number,
it mostly does if identical fermions are involved. As a consequence, exact
fermion Monte Carlo methods are limited to few-particle systems. In what
follows, we apply a Monte Carlo method which scales favorably and is
nevertheless exact.

The paper is organized as follows. In section II we will discuss the
principle of many-body diffusion (MBDF) \cite{BDL94,BDL95,BDL96,BDL97}. This
formalism is applied to derive random processes to numerically predict some
energy eigenvalues of the model Hamiltonian. A feasible implementation
scheme is discussed. Section III gives the outcome of our numerical
analysis, and a comparison to other work is made. Finally, section IV
concludes the present article.

\section{The many-body diffusion algorithm}

The model is approached with a recently developed quantum Monte Carlo
method, the many-body diffusion algorithm (MBDA) \cite{LBDL98,LBDL98b}. The
MBDA is based on the many-body diffusion formalism (MBDF) \cite
{BDL94,BDL95,BDL96,BDL97}. Here we will restrict ourselves to a brief
outline of the underlying concepts. In the MBDF, the propagator of $N$
interacting identical particles in $d$ spatial dimensions is written as a
Feynman-Kac functional \cite{KORZE} over a symmetrized process, i.e., as a
Euclidean-time path integral over the diffusion process of $N$ identical
free particles with superimposed potential-dependent exponential weights.
For a given irreducible symmetry representation $S$, the propagator 
\begin{equation}
K_{S}\left( \overline{r}_{f},\tau ;\overline{r}_{i}\right) =E_{\overline{r}%
_{i}}\left[ I_{\left( \overline{R}^{S}\left( \tau \right) =\overline{r}%
_{f}\right) }\exp \!\left( -\int\limits_{0}^{\tau }V\left( \overline{R}%
^{S}(\varsigma )\right) d\varsigma \right) \right]  \label{FKD}
\end{equation}
is hence represented as an average over all paths starting in $\overline{r}%
_{i}$, as indicated by the averaging index $E_{\overline{r}_{i}}$, and
ending a Euclidean time lapse $\tau $ later in $\overline{r}_{f}$, as
denoted by the indicator $I_{\left( \overline{R}\left( \tau \right) =%
\overline{r}_{f}\right) }$. The symmetry representation $S$ determines the
construction principle for the underlying $dN$-dimensional diffusion process 
$\left\{ \overline{R}^{S}\left( \tau \right) ;\tau \geq 0\right\} $.

In the MBDF, a detailed analysis has been performed of the diffusion process
of free identical particles, and the role of the potential symmetry has been
pointed out. It was found that for coordinate-symmetric potentials, i.e.,
potentials invariant under the permutation of the Cartesian particle
coordinates, and for certain irreducible symmetry representations $S$, the
total propagator separates into a sum of stochastically independent
sub-propagators. The importance of this coordinate-symmetry shows two-fold.
First, it allows to easily generalize the diffusion process of free
identical particles to the process of interacting identical particles.
Second, dealing with identical fermions, the sign problem is strictly
avoided by numerical procedures based on exclusively positive walkers. The
Feynman-Kac formulation (\ref{FKD}) indicates the relevance of the free
diffusion process. For any coordinate-symmetric potential $V(\bar{r})$, the
symmetry properties of the total diffusion process correspond to those of
the free one. Rather than interfering with the role of the potential,
different irreducible symmetry representations $S$ do completely determine
the structure of the free diffusion process. Correspondingly, one may in
principle approach various eigenstates of the system by formulating the free
diffusion process in the appropriate symmetry representations.

In \cite{BDL96}, the free density matrix of $N$ identical particles is
decomposed into corresponding one-dimensional $N$-particle density matrices.
In one dimension (1D), the introduction of an ordered $N$-particle state
space $\tilde{D}_{N}=\{x_{1}\geq x_{2}\geq \cdots \geq x_{N}\}$ projects the
density matrix on a mathematically well-defined expression for both bosons
and fermions. On that basis, the MBDF introduces the {\sl fermion diffusion
process} $\tilde{X}_{f}$ and the {\sl boson diffusion process} $\tilde{X}%
_{b} $ as a Brownian motion on the irreducible state space $\tilde{D}_{N}$
with absorbing respectively reflecting boundary conditions. The processes $%
\tilde{X}_{f}$ and $\tilde{X}_{b}$ serve as key ingredients for the
multi-dimensional formulation. With the decomposition into one-dimensional
fermion and boson diffusion processes, a scheme has been introduced to
sample the free density matrix for specific symmetric and antisymmetric
symmetry representations sign-problem-free.

The symmetry constraints specified in \cite{LBDL98,BDL96} rely on the
complete (anti-)symmetrization along the Cartesian coordinates. This scheme
addresses a particular excited fermion state. To illustrate this idea,
consider $N$ identical free fermions or bosons in two dimensions (2D) with
unit mass, for which the density matrix can be expressed as a determinant or
a permanent: 
\[
\rho _{\text{f}}\left( \overline{r}_{f},\tau ;\overline{r}_{i}\right)
=\det_{j,k=1,N}\left[ \rho \left( _{\text{d}}\vec{r}_{f}^{\text{ }j},\tau ;%
\vec{r}_{i}^{\text{ }k}\right) \right] \;\text{resp. }\rho _{\text{b}}\left( 
\overline{r}_{f},\tau ;\overline{r}_{i}\right) =%
%TCIMACRO{\limfunc{perm}}%
%BeginExpansion
\mathop{\rm perm}%
%EndExpansion
\limits_{j,k=1,N}\left[ \rho _{\text{d}}\left( \vec{r}_{f}^{\text{ }j},\tau ;%
\vec{r}_{i}^{\text{ }k}\right) \right] 
\]
where 
\[
\rho _{\text{d}}\left( \vec{r}_{f}^{\text{ }j},\tau ;\vec{r}_{i}^{\text{ }%
k}\right) =\frac{1}{2\pi \tau }\exp \!\left( -\frac{(\vec{r}_{f}^{\text{ }j}-%
\vec{r}_{i}^{\text{ }k})^{2}}{2\tau }\right) . 
\]
Complete particle (anti)symmetrization along the Cartesian directions then
leads to the following representation for $N$ two-dimensional identical
fermions 
\[
\rho _{\text{f}}\left( \overline{r}_{f},\tau ;\overline{r}_{i}\right)
=\det_{j,k=1,N}\left[ \rho _{\text{d}}\left( \vec{r}_{f}^{\text{ }j},\tau ;%
\vec{r}_{i}^{\text{ }k}\right) \right] =\rho _{\text{f}}\left( \overline{x}%
_{f},\tau ;\overline{x}_{i}\right) \rho _{\text{b}}\left( \overline{y}%
_{f},\tau ;\overline{y}_{i}\right) +\rho _{\text{b}}\left( \overline{x}%
_{f},\tau ;\overline{x}_{i}\right) \rho _{\text{f}}\left( \overline{y}%
_{f},\tau ;\overline{y}_{i}\right) . 
\]
Applying an analogous decomposition for three spatial dimensions, excited
state energies of up to 20 harmonically interacting identical fermion
oscillators have been efficiently simulated within a statistical accuracy of
about 0.1 percent \cite{LBDL98}. Quadratic particle interaction has also
been considered in the framework of quantum dots \cite{JOH91} but is not
studied here. By an extension of the many-body diffusion principle, it is
also possible to extract the corresponding ground-state energies. Again,
particle (anti)symmetrization along the Cartesian directions plays the
central role in that formulation. However, in contrast to \cite{LBDL98,BDL96}%
, not all particles are necessarily (anti)symmetrized in the same direction.
A detailed discussion of the underlying formalism is beyond the scope of the
present paper and will be published elsewhere.

The derivation of the sampled functional separates into two parts, the
decomposition of the corresponding free density matrix and symmetry
considerations on the potential. The underlying free diffusion principle is
best explained for the example of three identical free fermions in two
dimensions \cite{Openshell}. In this case, the infinite-time limit of the
free density matrix reads 
\begin{eqnarray}
\lim_{\tau \rightarrow \infty }\frac{\rho _{\text{f}}\left( \overline{r}%
_{f},\tau ;\overline{r}_{i}\right) }{\rho _{\text{d}}\left( \overline{r}%
_{f},\tau ;\overline{r}_{i}\right) e^{\overline{r}_{f}\cdot \overline{r}%
_{i}/\tau }} &=&\frac{1}{\tau ^{2}}\left| 
\begin{array}{lll}
1 & x_{1} & y_{1} \\ 
1 & x_{2} & y_{2} \\ 
1 & x_{3} & y_{3}
\end{array}
\right| \left| 
\begin{array}{ccc}
1 & x_{1}^{\prime } & y_{1}^{\prime } \\ 
1 & x_{2}^{\prime } & y_{2}^{\prime } \\ 
1 & x_{3}^{\prime } & y_{3}^{\prime }
\end{array}
\right| +O(\tau ^{-3})  \nonumber \\
&=&\frac{1}{\tau ^{2}}\left( 
\begin{array}{c}
\left[ (x_{1}-x_{2})(y_{2}-y_{3})\right] \left[ (x_{1}^{\prime
}-x_{2}^{\prime })(y_{2}^{\prime }-y_{3}^{\prime })\right] \\ 
+\left[ (x_{2}-x_{3})(y_{1}-y_{2})\right] \left[ (x_{2}^{\prime
}-x_{3}^{\prime })(y_{1}^{\prime }-y_{2}^{\prime })\right] \\ 
-\left[ (x_{1}-x_{2})(y_{2}-y_{3})\right] \left[ (x_{2}^{\prime
}-x_{3}^{\prime })(y_{1}^{\prime }-y_{2}^{\prime })\right] \\ 
-\left[ (x_{2}-x_{3})(y_{1}-y_{2})\right] \left[ (x_{1}^{\prime
}-x_{2}^{\prime })(y_{2}^{\prime }-y_{3}^{\prime })\right]
\end{array}
\right) +O(\tau ^{-3}).  \label{three}
\end{eqnarray}

As our concern is the generation of equilibrated samples to derive
properties of the (lowest available) eigenstate, the use of this asymptotic
limit is justified. Emphasis in our approach --as mostly in random-walk or
diffusion Monte Carlo approaches-- is thus on the long-term distribution
rather than on the equilibration process itself. Eq. (\ref{three}) involves
two interdependent stochastic processes. They both include a Brownian
motion, but distinguished by their respective positive domains $D_{1}$ and $%
D_{2}$, 
\[
D_{1}=\left\{ x_{1}\geq x_{2};\,y_{2}\geq y_{3}\right\} \text{ and }%
D_{2}=\left\{ x_{2}\geq x_{3};\,y_{1}\geq y_{2}\right\} \text{,} 
\]
and boundary conditions. Apart from an adapted distinguishable-particle
diffusion, a jump process must be realized to take into account the process
interdependencies \cite{LBDL98b}. In practice, during an evolution cycle, a
walker associated to a particular type of process might be assigned to the
other type of process according to the locally dependent process transition
rates. In the present case, in which we are interested in the derivation of
the ground-state energy, single-process evolution is sufficient for our
needs. Indeed, as long as the time decay rates of the individual processes
are identical, the energy eigenvalue predicted by any of the single
processes is the same as the one predicted from their combination.

Table 1 shows the symmetry configurations for the numerical simulation of
the ground-state energy of three and six two-dimensional non-interacting
spin-polarized harmonic fermions. The ground-state energy of three identical
fermions in 2D, e.g., can be simulated by anti-symmetrization of pairs of
Cartesian coordinates, namely $(x_{1},x_{2})$ and $(y_{2},y_{3})$.
Analogously, one might conglomerate the coordinates $(x_{2},x_{3})$ and $%
(y_{1},y_{2})$. For six identical fermions in 2D, e.g., one simultaneously
anti-symmetrizes $(x_{1},x_{2},x_{3})$, $(x_{4},x_{6})$, $%
(y_{3},y_{4},y_{5}) $ , and $(y_{2},y_{6})$, and so forth.

The advantage of the (generalized) many-body diffusion approach lies in the
efficient sampling of the required probability densities. As mentioned
above, symmetry and anti-symmetry in one spatial dimension can be achieved
by the definition of reflecting and absorbing boundary conditions on a
distinguishable-particle Brownian motion. Neither do we have to deal with
negative transition amplitudes, nor is our approach slowed down by sampling
determinants. A detailed description of the algorithmic realization has been
reported in \cite{LBDL98}. The Euclidean-time evolution according to $%
K_{S}\left( \overline{r}_{f},\tau ;\overline{r}_{i}\right) $ is simulated in
sufficiently small time steps $\epsilon $ by the repeated application of the
following two-step procedure: a) given ($2N$-dimensional) initial system
configurations $\overline{r}_{i}$ sample final ones $\overline{r}_{f}(\tau
+\epsilon )=\overline{r}_{i}(\tau )+\delta \overline{r}(\epsilon )$, where $%
\delta \overline{r}(\epsilon )$ is randomly drawn according to the free
identical-particle propagator, and b) apply the potential-dependent weights $%
\exp \left[ -%
%TCIMACRO{\tint}%
%BeginExpansion
\textstyle\int%
%EndExpansion
\nolimits_{\tau }^{\tau +\epsilon }V\left( \overline{R}^{S}(\varsigma
)\right) d\varsigma \right] $ randomly in a branching and killing procedure.
The determination of the weights $\exp \left[ -%
%TCIMACRO{\tint}%
%BeginExpansion
\textstyle\int%
%EndExpansion
\nolimits_{\tau }^{\tau +\epsilon }V\left( \overline{R}^{S}(\varsigma
)\right) d\varsigma \right] $ in principle requires infinitesimal time steps 
$\epsilon \rightarrow 0$. Due to limited computer performance, however, this
procedure is not practical, and reliable approximation schemes must be
provided. In the present case of harmonic confinement and repulsive Coulomb
interaction, the use of the Suzuki-Trotter weights $\exp \left\{ -\frac{1}{2}%
\left[ V\left( \overline{R}^{S}(\tau )\right) +V\left( \overline{R}^{S}(\tau
+\epsilon )\right) \right] \right\} $ is satisfactory for realistic time
steps of 0.001/Hartree (H). The essential requirement for the efficient
approach of many-body systems with the outlined free diffusion construction
principles is the coordinate-symmetry of the potential involved. This
condition holds for both the confinement and the interaction potential of
our model 
\begin{equation}
\forall \left( i,j\in \{1,2,3\}\vee \{4,5,6\}\right) :V_{\text{int}}(\bar{r}%
)=V_{\text{int}}(\hat{P}_{x}^{i,j}\bar{r})=V_{\text{int}}(\hat{P}_{y}^{i,j}%
\bar{r})\wedge V_{\text{c}}(\bar{r})=V_{\text{c}}(\hat{P}_{x}^{i,j}\bar{r}%
)=V_{\text{c}}(\hat{P}_{y}^{i,j}\bar{r}).  \label{coordsym}
\end{equation}
In (\ref{coordsym}), the operators $\hat{P}_{x}^{i,j}$ and $\hat{P}%
_{y}^{i,j} $ interchange the $i$th and the $j$th $x$- respectively $y$%
-coordinates.

It should be emphasized that this invariance of the Hamiltonian under the
interchange of the $x$ and $y$ coordinates of any two particles no longer
applies if the confinement potential is replaced for instance by a Coulomb
potential. The potential then generates transition rates between different
types of walkers, as discussed in detail in \cite{LBDL98b}. Although the
principle of sign-problem free diffusion remains valid for such systems, the
detailed analytical analysis of the process interdependencies becomes
cumbersome. With our present approach of Carthesian decomposition, more than
six electrons become almost intractable in practice. New purely numerical
techniques are currently under development to perform the required symmetry
decompositions, avoiding the tedious and unpractical analytical bookkeeping
of the Carthesian decompositions for the different types of walkers.
Preliminary studies reveal that this numerical analysis requires in general
the evaluation of $N\times N$ determinants relating the initial and final
positions of the walkers. This $N^{3}$ cost in computation time is
presumably overcompensated by a factor $1/N!$ due to the reduction of the
state space. Within the present status of our approach however, closed shell
systems with potentials satisfying the symmetry condition (\ref{coordsym})
are tractable for as many as 20 electrons.

\section{Results and Discussion}

Table 2 gives the ground-state energy estimates predicted for closed-shell
systems of six, twelve and twenty unpolarized electrons as a function of $%
\gamma ^{1/3}$. The energies obtained are supposed to be exact within the
numerically estimated statistical error. In the limit of zero electron
Coulomb repulsion, $\gamma \rightarrow 0$, (1) is identical to a system of
non-interacting fermion oscillators. The arrangement of single-particle
harmonic oscillator solutions into a Slater determinant then induces the
energy limits $E_{0}^{\gamma \rightarrow 0}=10$ H, $E_{0}^{\gamma
\rightarrow 0}=28$ H and $E_{0}^{\gamma \rightarrow 0}=60$ H for six, twelve
and twenty electrons, respectively, whereas the neglect of spin statistics
would yield $E_{\text{dist}}^{\gamma \rightarrow 0}=N$ H, with $N$
indicating the particle number. In the opposite limit, for infinitely large
electron-electron repulsion, $\gamma \rightarrow \infty $, the electrons
behave classically and one expects them to arrange in the form of a Wigner
lattice \cite{BED94}. With increasing $\gamma $, the average
electron-electron distance grows, and the influence of spin-statistics
weakens. Accordingly, as $\gamma $ grows, the energy gap between different
excited states is expected to decrease substantially. This physical behavior
is recovered by our numerical data. The limit of zero electron-electron
repulsion is accurately simulated, and a smooth transition to high $\gamma $
is found. A comparison of the numerical energy eigenvalues and the energy of
distinguishable particles (see Table 3) indeed indicates the irrelevance of
quantum statistics as $\gamma \rightarrow \infty $.

It proves instructive to compare our energy estimates for the ground state
of the unpolarized electron systems with the Pad\'{e} approximants reported
in ref. \cite{GON97b}. The relative deviation of the Pad\'{e} approximants
with respect to our numerical estimates $E_{0}$ for the case of six
unpolarized electrons are also studied. For both zero and very large $\gamma 
$, the Pad\'{e} approximates $P_{3,2}(\gamma )$ and $P_{4,3}(\gamma )$ match
our prediction. For intermediate $\gamma $, the Pad\'{e} approximants $%
P_{4,3}(\gamma )$ for the ground state energy introduce a systematic error
of up to almost 4 per cent. A comparison of our numerical estimates for the
twelve and the twenty-particle system with the corresponding Pad\'{e}
approximates \cite{GON97b} indicates an analogous qualitative picture. The
extreme relative deviations are of the same order. The lack of an accurate
description of the system (1) for intermediate regions of the
electron-electron-repulsion parameter $\gamma $ is typical for a variety of
analytical approximations.

\section{Conclusions}

In the present paper, we apply a generalization of the recently reported
many-body diffusion formalism to a system of interacting electrons in a
parabolic confinement. The method is illustrated explicitly for closed-shell
configurations of six, twelve and twenty unpolarized electrons for which the
ground-state energy is numerically predicted. The algorithm proceeds without
the use of analytical approximations. Apart from a small but controllable
systematic error, the energy values are numerically exact within a computed
statistical error of a few per mil. The feasibility of our approach is
indicated by the comparison of our ground-state energy estimates with the
corresponding Pad\'{e} approximants calculated by A. Gonzales et al. \cite
{GON97b}. Analogous results \cite{Egger99} were obtained by another Monte
Carlo technique \cite{MAK98,MAK99} that strongly reduces the noise due to
the sign problem.

Regarding the methodological aspect, this work introduces an algorithm which
allows to efficiently simulate the ground-state energy of closed-shell
configurations of electrons exposed to coordinate-symmetric potentials. The
scheme strictly avoids the fermion sign problem \cite{SCH84,CEP96} by the
definition of a Brownian motion on a state space with the appropriate
boundary conditions. The resulting random process can be realized by stable
diffusion of purely positive walkers.

The formulation of a sign-problem-free algorithm for the quantum-dot model
(1) is remarkable, since almost all quantum Monte Carlo algorithms face
massive difficulties with the general description of significantly
correlated continuous quantum systems. The reason for this is the fermion
sign problem, which generally thwarts reliable stochastic many-fermion
treatments. Although potentially exact, the transient estimation of
eigen-energies experiences a serious inefficiency due to an exponential
decreasing signal-to-noise ratio. On the other hand, even the use of very
accurate and efficiently implemented trial wave functions in diffusion or
Green's function Monte Carlo variants \cite{FIXNO} introduces considerable
systematic errors due to the assumption of generally incorrect nodal
surfaces. The MBDA avoids these problems by the definition of a diffusion
process on a state space with absorbing and/or reflecting boundaries. Its
derivation is based on a symmetry analysis of the Hamiltonian only.

\label{0acknowledgments}\acknowledgments
This work is performed within the framework of the FWO\ projects No.
1.5.729.94, 1.5.545.98, G.0287.95, G.0071.98 and WO.073.94N
(Wetenschappelijke Onderzoeksgemeenschap van het FWO over ``Laagdimensionele
Systemen'', Scientific Research Community of the FWO on ``Low-Dimensional
Systems''), the ``Interuniversitaire Attractiepolen -- Belgische Staat,
Diensten van de Eerste Minister -- Wetenschappelijke, Technische en
Culturele Aangelegenheden'', and in the framework of the BOF\ NOI 1997
projects of the Universiteit Antwerpen. One of the authors (F.B.)
acknowledges the FWO for financial support.

%TCIMACRO{
%\TeXButton{B}{\begin{table}[tbp]%
%} }%
%BeginExpansion
\begin{table}[tbp]%
%
%EndExpansion
\begin{tabular}{l|l|}
N & antisymmetric constellation \\ \hline
$3$ & ($x_{1},x_{2}$); ($y_{2},y_{3}$) \\ \hline
$6$ & ($x_{1},x_{2},x_{3}$); ($x_{4,}x_{6}$); ($y_{3},y_{4},y_{5}$); ($%
y_{2,}y_{6}$) \\ \hline
$10$ & 
\begin{tabular}{c}
($y_{1},y_{2},y_{3},y_{4}$); ($y_{5},y_{8},y_{9}$); ($y_{6},y_{10}$) \\ 
($x_{4},x_{5},x_{6},x_{7}$); ($x_{3},x_{8},x_{10}$); ($x_{2},x_{9}$)
\end{tabular}
\\ \hline
\end{tabular}
\caption{List of configurations used for the ground states of N identical 2D fermions.
\label{key}} 
%TCIMACRO{
%\TeXButton{E}{\end{table}%
%} }%
%BeginExpansion
\end{table}%
%
%EndExpansion
\bigskip

%TCIMACRO{
%\TeXButton{B}{\begin{table}[tbp] \centering%
%} }%
%BeginExpansion
\begin{table}[tbp] \centering%
%
%EndExpansion
\begin{tabular}{c|cccccccc}
\hline
${\bf \gamma }^{1/3}$ & 0.2 & 0.4 & 0.6 & 0.8 & 1.0 & 1.2 & 1.4 & 1.6 \\ 
\hline
$E_{0}(3\!\!\uparrow ,3\!\!\downarrow )$ & 10.086(4) & 10.804(3) & 12.632(4)
& 15.806(4) & 20.367(3) & 26.158(2) & 33.084(4) & 41.115(5) \\ 
$E_{0}(6\!\!\uparrow ,6\!\!\downarrow )$ & 28.28(4) & 31.00(3) & 37.91(2) & 
49.62(2) & 66.45(1) & 87.88(2) & 113.52(2) & 143.45(2) \\ 
$E_{0}(10\!\!\uparrow ,10\!\!\downarrow )$ & 60.99(7) & 67.72(9) & 85.24(3)
& 114.98(8) & 157.97(4) & 211.70(8) & 276.72(7) & 351.72(6) \\ \hline
\end{tabular}
\caption{List of energy eigenvalues (a.u.) for model (1) obtained by the MBDA.
\label{key2}} 
%TCIMACRO{
%\TeXButton{E}{\end{table}%
%}}%
%BeginExpansion
\end{table}%
%
%EndExpansion

%TCIMACRO{
%\TeXButton{B}{\begin{table}[tbp]%
%} }%
%BeginExpansion
\begin{table}[tbp]%
%
%EndExpansion
\begin{tabular}{c|cccccc}
\hline
& $E_{0}(3\!\!\uparrow ,3\!\!\downarrow )$ & $E_{\text{dist}}(3\!\!\uparrow
,3\!\!\downarrow )$ & $E_{0}(6\!\!\uparrow ,6\!\!\downarrow )$ & $E_{\text{%
dist}}(6\!\!\uparrow ,6\!\!\downarrow )$ & $E_{0}(10\!\!\uparrow
,10\!\!\downarrow )$ & $E_{\text{dist}}(10\!\!\uparrow ,10\!\!\downarrow )$
\\ \hline
${\bf \gamma }^{1/3}=2$ & 60.385(5) & 60.12(1) & 215.21(1) & 214.30(3) & 
533.56(3) & 531.44(4) \\ \hline
\end{tabular}
\caption{List of energy eigenvalues (a.u.) for model (1) obtained by the MBDA.
\label{key3}} 
%TCIMACRO{
%\TeXButton{E}{\end{table}%
%}}%
%BeginExpansion
\end{table}%
%
%EndExpansion


\begin{references}
\bibitem{Openshell}  We limit the discussion to closed shell configurations,
not because our method is not applicable for open shells, but for reasons of
relative algebraic simplicity. For open shell systems the degeneracy of the
ground state induces a state space with linear combinations of possible
initial and final states . This degeneracy poses no problems of principle,
but the resulting conditions become quite involved. E.g., for 4 particles in
two dimensions the degeneracy is threefold, and the subprocesses and their
boundary conditions have to be determined from the following infinite-time
limit of the free density matrix \newline
\[
\lim_{\tau \rightarrow \infty }\frac{\rho _{\text{f}}\left( \overline{r}%
_{f},\tau ;\overline{r}_{i}\right) }{\rho _{\text{d}}\left( \overline{r}%
_{f},\tau ;\overline{r}_{i}\right) e^{\overline{r}_{f}\cdot \overline{r}%
_{i}/\tau }}=\frac{1}{2\tau ^{4}}\left( 
\begin{array}{c}
\left| 
\begin{array}{cccc}
1 & x_{1}^{\prime } & y_{1}^{\prime } & (x_{1}^{\prime })^{2} \\ 
1 & x_{2}^{\prime } & y_{2}^{\prime } & (x_{2}^{\prime })^{2} \\ 
1 & x_{3}^{\prime } & y_{3}^{\prime } & (x_{3}^{\prime })^{2} \\ 
1 & x_{4}^{\prime } & y_{4}^{\prime } & (x_{4}^{\prime })^{2}
\end{array}
\right| \left| 
\begin{array}{llll}
1 & x_{1} & y_{1} & x_{1}^{2} \\ 
1 & x_{2} & y_{2} & x_{2}^{2} \\ 
1 & x_{3} & y_{3} & x_{3}^{2} \\ 
1 & x_{4} & y_{4} & x_{4}^{2}
\end{array}
\right| \\ 
+\left| 
\begin{array}{cccc}
1 & x_{1}^{\prime } & y_{1}^{\prime } & (y_{1}^{\prime })^{2} \\ 
1 & x_{2}^{\prime } & y_{2}^{\prime } & (y_{2}^{\prime })^{2} \\ 
1 & x_{3}^{\prime } & y_{3}^{\prime } & (y_{3}^{\prime })^{2} \\ 
1 & x_{4}^{\prime } & y_{4}^{\prime } & (y_{4}^{\prime })^{2}
\end{array}
\allowbreak \right| \left| 
\begin{array}{llll}
1 & x_{1} & y_{1} & y_{1}^{2} \\ 
1 & x_{2} & y_{2} & y_{2}^{2} \\ 
1 & x_{3} & y_{3} & y_{3}^{2} \\ 
1 & x_{4} & y_{4} & y_{4}^{2}
\end{array}
\right| \\ 
+2\left| 
\begin{array}{cccc}
1 & x_{1}^{\prime } & y_{1}^{\prime } & x_{1}^{\prime }y_{1}^{\prime } \\ 
1 & x_{2}^{\prime } & y_{2}^{\prime } & x_{2}^{\prime }y_{2}^{\prime } \\ 
1 & x_{3}^{\prime } & y_{3}^{\prime } & x_{3}^{\prime }y_{3}^{\prime } \\ 
1 & x_{4}^{\prime } & y_{4}^{\prime } & x_{4}^{\prime }y_{4}^{\prime }
\end{array}
\right| \left| 
\begin{array}{llll}
1 & x_{1} & y_{1} & x_{1}y_{1} \\ 
1 & x_{2} & y_{2} & x_{2}y_{2} \\ 
1 & x_{3} & y_{3} & x_{3}y_{3} \\ 
1 & x_{4} & y_{4} & x_{4}y_{4}
\end{array}
\right|
\end{array}
\right) . 
\]
\newline
The complexity increases with increasing degeneracy of the fermion ground
state, and the bookkeeping of the carthesian decompositions soon becomes
very complicated. More appropriate methods for handling open shells are
under study.
\end{references}
\end{document}